\documentclass[draft]{agujournal2019}
\usepackage{url} 
\usepackage{lineno}
\usepackage[inline]{trackchanges} 
\usepackage{soul}
\usepackage{ragged2e}
\justifying
\usepackage{graphicx} 
\usepackage[utf8]{inputenc}
\journalname{Geophysical Research Letters}

\begin{document}
%
%


\title{Low frequency radio observations of the `quiet' corona during the descending phase of sunspot cycle 24}

%
%




\authors{R. Ramesh\affil{1}, A. Kumari\affil{1,4}, C. Kathiravan\affil{1}, 
D. Ketaki\affil{1,2}, M. Rajesh\affil{1}, and M. Vrunda\affil{1,3}}


\affiliation{1}{Indian Institute of Astrophysics, Koramangala 2nd Block, Bangalore, Karnataka, India - 560034}
\affiliation{2}{Sir Parashurambhau College, Pune, Maharashtra, India - 411 030}
\affiliation{3}{Skolkovo Institute of Science and Technology, 3, Nobel Street, Moscow - 121205, Russian Federation}
\affiliation{4}{Department of Physics, University of Helsinki, P.O. Box 64, FI-00014 Helsinki, Finland}





\correspondingauthor{R. Ramesh}{ramesh@iiap.res.in}




\begin{keypoints}
\item We investigated the equatorial diameter of `quiet' solar corona at low radio frequencies during the descending phase of sunspot cycle 24.
\item Localized sources of emission in the radio images were identified and removed using iterative multi-Gaussian curve fitting technique.
\item The minimum radius of the `quiet' solar corona at a typical frequency like 53 MHz was found to be ${\approx}1.16\rm R_{\odot}$.
\end{keypoints}

%
%

%
%


\begin{abstract}
We carried out a statistical study of the `quiet' solar corona during the descending phase of the sunspot cycle 24 (i.e. 2015 January - 2019 May) using data obtained with the Gauribidanur RAdioheliograPH (GRAPH) at 53 MHz and 80 MHz simultaneously. Our results show that the equatorial (east-west) 
diameters of the solar corona at the above two frequencies shrunk steadily.
The decrease was found to be due to a gradual reduction in the coronal electron 
density ($N_{e}$).
Independent estimates of $N_{e}$ in the equatorial region of the `background' corona using white-light coronagraph observations indicate a decline consistent with our findings. 
\end{abstract}

\section{Introduction}

One of the well established result from the white-light observations during total solar eclipses and with coronagraphs is that the shape/size and electron density of the corona varies with the sunspot cycle.  This is because of the close correspondence between the coronal brightness and the electron density which in turn is an indicator of the coronal magnetic field (see for e.g. \citeA{MacQueen2001}. These studies were primarily about the `global' corona which included structures like the coronal streamers also. We wanted to investigate the `quiet' solar corona (i.e., the corona distinct from emission due to transient and long-lasting discrete sources) at radio frequencies during the sunspot cycle 24 since it could provide independent results related to the latter. 
Observations of thermal bremsstrahlung radio emission from the `quiet' solar corona at frequencies $<$100 MHz are expected to be particularly useful in this connection since the radiation originates fully in the corona. The corona above the visible solar disk and off the solar limb can be simultaneously imaged. 
Hence the present work.

\section{Observations}

The radio observations were carried out using the facilities operated by the Indian Institute of Astrophysics (IIA) in the Gauribidanur observatory  (\citeA{Ramesh2011a,Ramesh2014}; https://www.iiap.res.in/?q=centers/radio). 
Two-dimensional radio images obtained with the \textit{Gauribidanur RAdioheliograPH} (GRAPH; \citeA{Ramesh1998,Ramesh1999a,Ramesh2006b}) at 80 MHz and 53 MHz during the period 2015 January - 2019 May were used. The GRAPH is a T-shaped long radio interferometer array and it has an angular resolution (`beam') of 
${\approx}4^{\prime}{\times}6^{\prime}$ (R.A.\,$\times$\,decl.)
and ${\approx}6^{\prime}{\times}9^{\prime}$
at the above two frequencies, respectively.
The integration time is $\approx$250 msec and the observing bandwidth is $\approx$2 MHz. Observations are carried out everyday for $\approx$2.5 hr on either side of  
the transit of the Sun over the local meridian in Gauribidanur. However for the present work, we limited ourselves to the data obtained for a period of $\pm15$ min around the transit. Since the source zenith angle will be the least during that time, any possible refraction effects due to the ionosphere are generally minimal \cite{Stewart1982,Jacobson1991,Mercier1996}. Note that the local latitude of Gauribidanur is $\approx +14^{\circ}$ N. This is within the decl. range of $-23^{\circ}$ S to $+23^{\circ}$ N over which the Sun moves back and forth every year. So, even when the Sun is at $-23^{\circ}$ decl., its elevation for GRAPH would be 
high (${\approx}53^{\circ}$). For radio spectral data, we used observations with the \textit{Gauribidanur LOw-frequency Solar Spectrograph} (GLOSS; \citeA{Ebenezer2001,Ebenezer2007,Kishore2014,Hariharan2016b}), \textit{Gauribidanur RAdio Spectro-Polarimeter} (GRASP; \citeA{Sasi2013a,Kishore2015,Hariharan2015}), 
and \textit{e-CALLISTO} \cite{Monstein2007,Benz2009}. We also used data obtained with the \textit{Gauribidanur Radio Interferometric Polarimeter} (GRIP; \citeA{Ramesh2008}). The combined use of the aforementioned imaging, spectral, and polarimetric data help to understand the radio signatures associated with the corresponding solar activity in a better 
manner (see for e.g. \citeA{Sasikumar2013b}). The optical data were obtained in white-light with the COR1 coronagraph of the \textit{Sun-Earth Connection Coronal and Heliospheric Investigation} (SECCHI; \citeA{Howard2008}) on board the \textit{Solar Terrestrial Relationship Observatory-A} (STEREO-A, https://cor1.gsfc.nasa.gov/).

The GRAPH data were carefully selected such that: (i) no H$\alpha$ and/or X-ray flares, coronal mass ejections (CMEs), and short/long duration non-thermal radio burst activities were reported during our observing period as well as for $\approx$30 min before and after (ftp://ftp.swpc.noaa.gov/pub/warehouse; http://sidc.oma.be/cactus/).
In addition to the above, we verified the presence/absence of radio burst activities using data obtained with GLOSS, GRASP and e-CALLISTO also. Observations with the GRIP 
were particularly useful to verify the presence/absence of discrete sources of weak non-thermal emission like noise storm continuum on any given day via their circularly polarized radio emission and accordingly select the GRAPH data (see for e.g. \citeA{Ramesh2011b,Mugundhan2018b,McCauley2019}). Note that observations with a correlation radio interferometer with mutually perpendicular linearly polarized antennas as in GRIP are sensitive for circular polarization measurements (see for e.g. \citeA{Thompson2001}). 
With all the aforementioned checks, we unambiguously identified GRAPH observations of `quiet' Sun for 494 days;
(ii) the spectral index between 53 MHz and 80 MHz is expected to be in the range 
$\approx$ +1.6 to +3.9 for thermal emission from the solar corona  \cite{Erickson1977,Subramanian1988,Ramesh2000d,Ramesh2006a,Oberoi2017}. 
But we considered only observations with spectral index $+2.0\pm0.2$ 
from the aforementioned 494 days of observations 
to minimize confusion due to possible contribution of non-thermal emission and/or temperature gradients in the corona (see for e.g. \citeA{Subramanian1988}).
This resulted in 382 days of concurrent data at the two frequencies for further analysis.

\section{The Method}

\begin{figure}
\noindent\includegraphics[width=\textwidth]{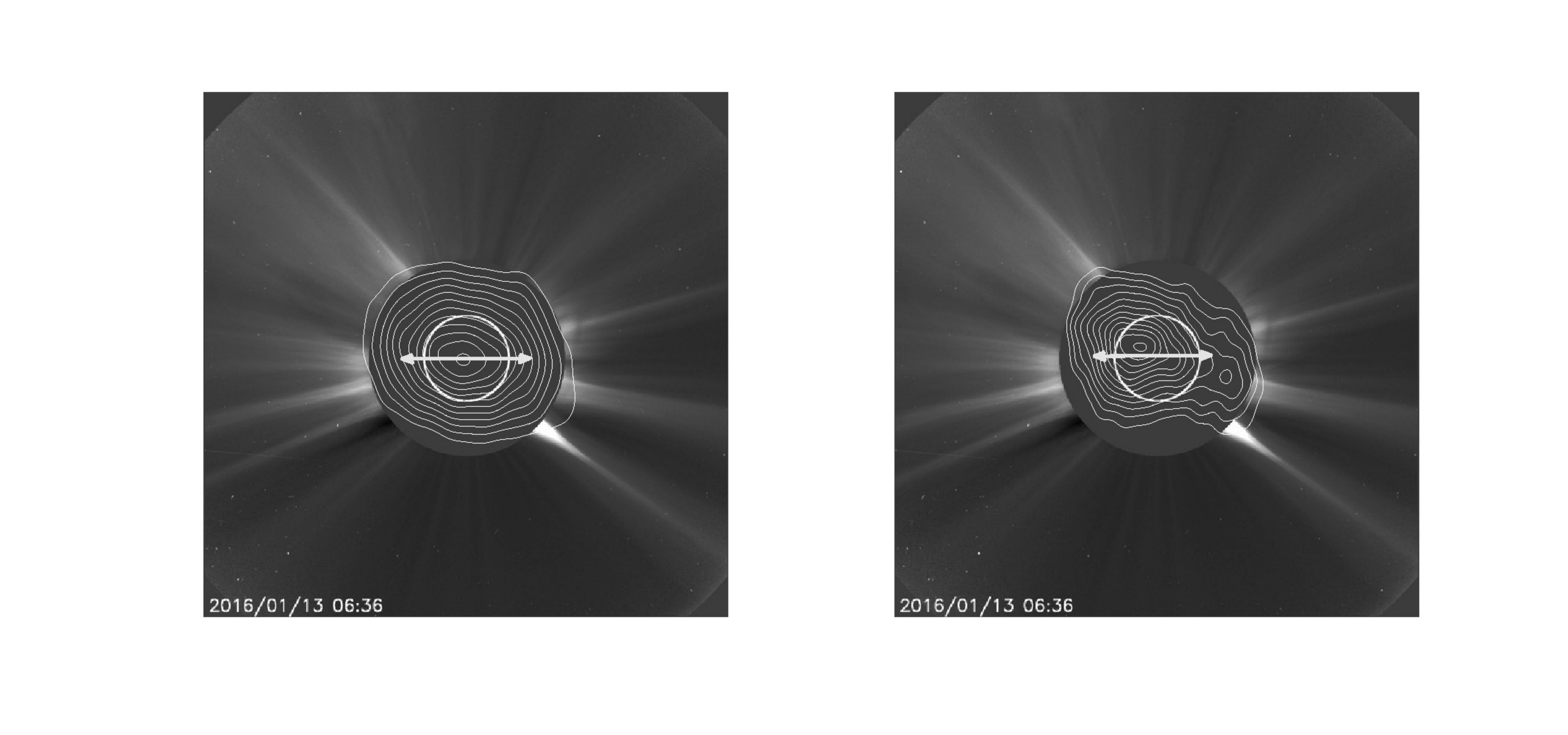}
\caption{GRAPH images of the solar corona obtained on 2016 January 13
at 53 MHz and 80 MHz superposed on the SOHO/LASCO-C2 whitelight image obtained on the same day around the same time. Solar north is straight up and east is to the left. 
The `white' colour open circles represent the size of the solar photosphere. 
The concentric, filled `black' colour circles indicate the occulting disk of the coronagraph. Its radius is ${\approx}2.2\rm R_{\odot}$. 
The horizontal `white' colour lines indicate the equatorial diameter of the `quiet' solar corona at the respective frequencies estimated using the method described in Section 3 of the text 
(see Figure 2).}
\label{figure1}
\end{figure}

We wanted to investigate the size of the `quiet' solar corona by measuring its 
equatorial diameter 
at different epochs 
using two-dimensional (2D) radio imaging data obtained with the GRAPH. Note that we limited ourselves to diameter measurements since it is a directly observable quantity. The equatorial diameter was particularly chosen for the study due to the comparatively better and declination independent angular resolution of the GRAPH in the east-west direction. Further the polar diameter are generally biased by the equatorward extension of the polar coronal holes during the descending phase of a sunspot cycle \cite{Hundhausen1981}.
To obtain the equatorial diameter, we determined the one-dimensional (1D) 
brightness distribution averaged over an angular width of ${\approx}5^{\prime}$
centered on the solar 
equator (i.e. where the decl. co-ordinate in the y-axis is equal to the decl. of the Sun on that day) in the 2D images obtained with the GRAPH (see for e.g. Figure \ref{figure1}). Since the above angular width is small, the `critical' plasma level (corresponding to any particular observing frequency) at different locations within  the enclosed coronal regions can be assumed to be at the same heliocentric distance. 
Each 1D brightness distribution obtained as mentioned above was reproduced using an iterative multi-Gaussian least squares curve fitting technique as described in \citeA{Ramesh2006a}. The minimum width of each Gaussian profile was limited to $1^{\prime}$, the smallest source size reported at frequencies $<$ 100 MHz \cite{Ramesh1999b,Ramesh2001b,Ramesh2000c,Kathiravan2011,Ramesh2012b,
Mugundhan2016,Mugundhan2018a}. 
The typical discrete thermal sources in the above frequency range are mostly the radio counterparts of large scale structures like the coronal streamers and/or coronal holes.  
The source sizes are ${\approx}8^{\prime}$ at 73.8 MHz \cite{Kundu1977,Lantos1987,Schmahl1994}.
The above size measurements indicate that the angular resolutions of the GRAPH at 53 MHz and 80 MHz are reasonably sufficient to identify discrete sources of thermal radio emission at the respective frequencies in the present case. This is true even for weak noise storm continuum sources, the other suspected candidate of slowly varying emission at frequencies $<$100 MHz \cite{Lantos1999b}, since their angular size is 
${\approx}10^{\prime}$ at 73.8 MHz \cite{Kundu1990}. 

\begin{figure}
\noindent\includegraphics[width=\textwidth]{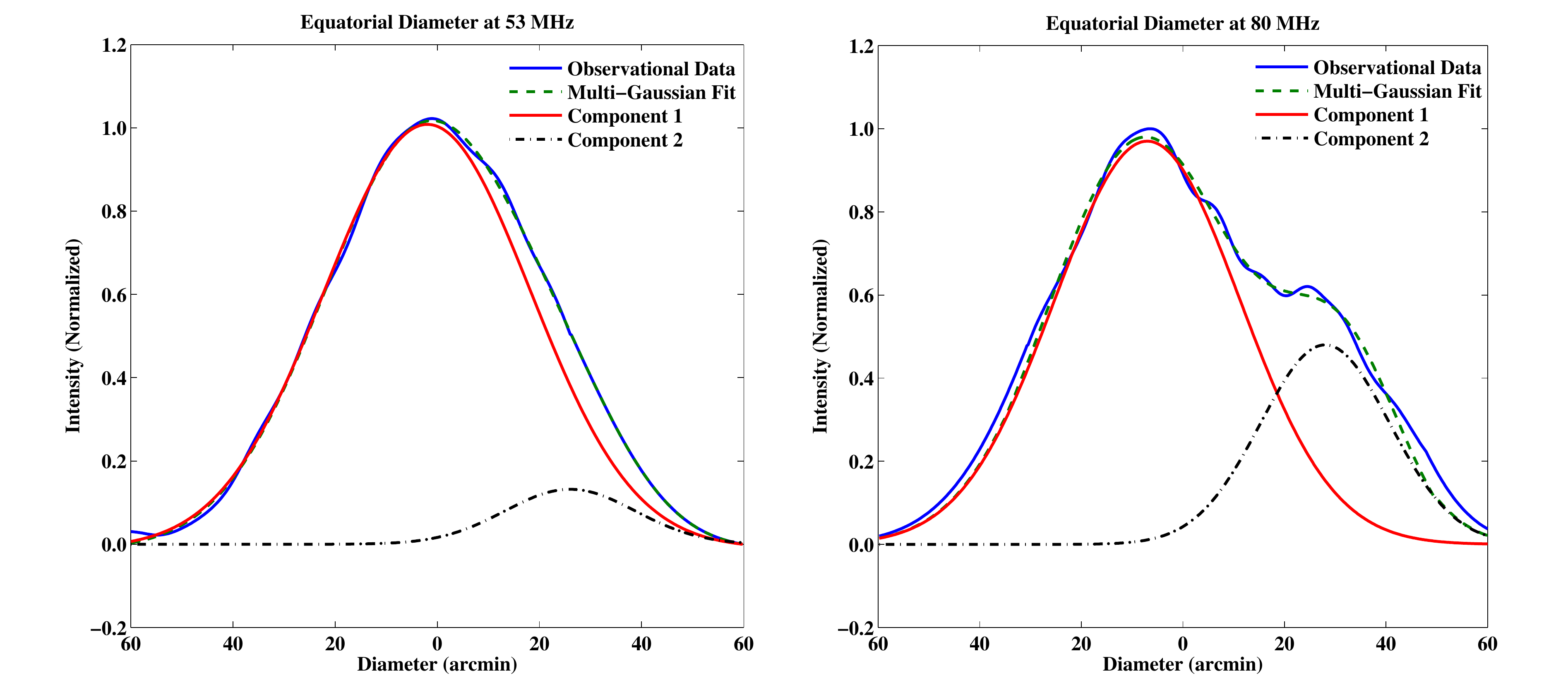}
\caption{1D equatorial brightness distribution obtained from the GRAPH radioheliograms at 53 MHz and 80 MHz (see Figure \ref{figure1}) based on the methodology described in Section 3. Two Gaussian profiles, red colour (solid line) and black colour (with dot-dash symbols), were used to match the observations (blue colour profile). The green colour profile (with dash symbols) is the sum of the aforementioned two Gaussian profiles used for the fit. The Chi-square fit errors are $<$1\% at both 53 MHz and 80 MHz.}
\label{figure2}
\end{figure}

We also imposed the following conditions to consider the data obtained at a particular epoch for further analysis: (i) the fits for the 53 MHz and 80 MHz observations must have one Gaussian profile each with width ${>}32^{\prime}$ 
and the width of the such a profile in the fit for the 53 MHz observations should be larger than the corresponding 80 MHz profile on the same day;
(ii) the discrete solar radio sources identified from the fit 
should be present at both the frequencies; (iii) the sum of the amplitudes of the Gaussian profiles used to fit the observed 1D brightness distribution at any given position on the latter should be nearly equal to the observed amplitude there. These conditions further reduced our data set to 336 days (see Section 2).
The iterative fit procedure was carried out till the difference in the amplitudes between the fit and the observations is $=<$0.05 SFU 
(1 SFU=$\rm 10^{-22}Wm^{-2}Hz^{-1}$), the minimum detectable flux density with GRAPH.
In addition to the Gaussian of width ${>}32^{\prime}$,
we used a maximum of four additional Gaussians to get the best fit on any given day at each of the two frequencies. 

The Gaussian profiles that satisfied the criteria (i) mentioned above were considered to represent the `background' corona, and their actual widths in the respective fits were considered to be the equatorial diameters at the corresponding frequencies for the observations on a particular day.
Assuming Gaussian shape for the GRAPH beam, we eliminated its contribution to the diameter estimate using the
formula $\theta_{s}{=}\sqrt{\left(\theta_{e}^{2}-\theta_{b}^{2}\right)}$, 
where $\theta_{s}$ is the source size without beam contribution, $\theta_{e}$ is the estimated source size as described above, and $\theta_{b}$ is the beam size.
An example of the diameter measurements using the above method is shown in Figure \ref{figure2}. There is excellent correspondence between the observations and the multi-Gaussian fit. Discrete sources close to the limb that are not so clearly discernible in the 2D radio images (for e.g. the source near the west limb in the 80 MHz image in Figure \ref{figure1} is less prominent in the 53 MHz image) are also effectively identified and removed (see Figure \ref{figure2}). The incomplete removal of such sources could lead to a larger equatorial diameter. 

 A comparison of the radio and whitelight images in Figure \ref{figure1} indicate that the radio contours are elongated in the east-west direction. This is predominantly due to enhanced radio emission associated with the multiple, closely spaced bright whitelight structures seen above the occulting disk of the coronagraph in the east and west directions. The contours appear smoothed because of the differences in angular resolution between radio and optical observations. An inspection of the near-Sun  
$({\approx}$1.05-3$\rm R_{\odot}$) whitelight coronagraph images (https://download.hao.ucar.edu/2016/01/12/20160113{\_}020824{\_}kcor{\_}l1{\_}cropped.gif) is in support of the above arguments. Note that SOHO/LASCO-C2 observations correspond to heliocentric distances 
$\rm{>}2.2R_{\odot}$ and hence the structures are seen distinctly. 
The differences between the 53 MHz and 80 MHz images themselves are chiefly due to the comparitively higher angular resolution of the GRAPH at 80 MHz and scattering of radio waves by density inhomogeneities in the solar corona \cite{Aubier1971,Thejappa1992,Ramesh2000a}. It was shown by 
\citeA{Thejappa2008} that positional displacements due to scattering could be 
${\sim}5^{\prime}$ at 50 MHz compared to 73.8 MHz. Earlier, \citeA{Sastry1994} had reported observations of 
${\sim}10^{\prime}$ shift between the centroids of the `quiet' Sun and the optical Sun. The `resolved' discrete radio source near the west limb in the 80 MHz image as against the 53 MHz image, and the separation between the centroids of the `quiet' Sun in the 53 MHz and 80 MHz images are consistent with the above arguments.

\begin{figure}
\noindent\includegraphics[width=\textwidth]{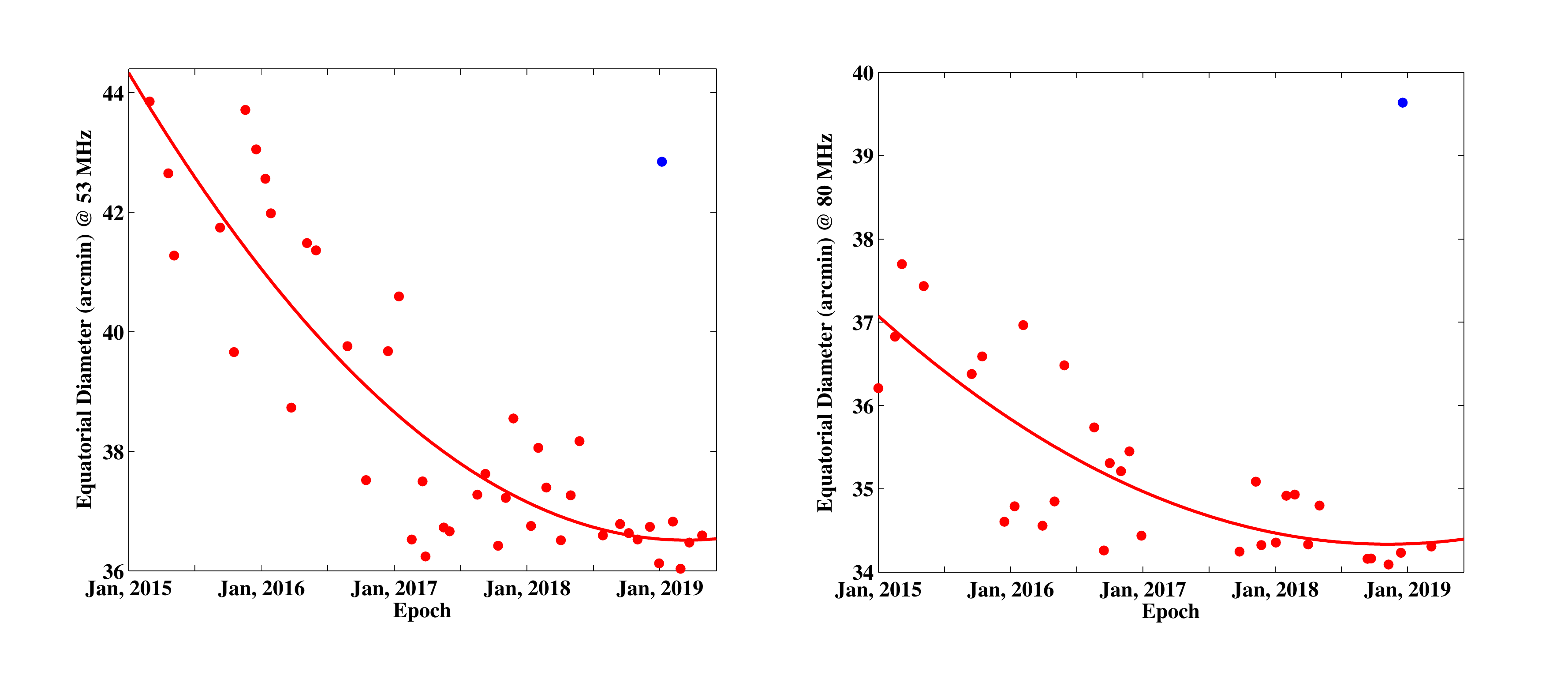}
\caption{The data points in red colour are the monthly minimum values in the equatorial 
diameters of the solar corona at 53 MHz (left panel) and 80 MHz (right panel) during the period 2015 January - 2019 May. The solid lines in red colour are the quadratic least squares fit 
to the data points. The isolated blue colour data point indicates the monthly maximum value in the equatorial diameter during 2019 January. It is significantly larger than the corresponding monthly minimum value even during the minimum period of sunspot cycle 24 like the aforementioned epoch.}
\label{figure3}
\end{figure}

Figure \ref{figure3} shows the monthly minimum values in the equatorial diameters of the solar corona at 53 MHz and 80 MHz during the period 2015 January - 2019 May estimated using the above methodology. The gap in the data points during 2017, particularly at 80 MHz, is due to the use of GRAPH for observations of background cosmic radio sources that are occulted by the solar corona at the above frequency (see for e.g. \citeA{Slee1961,Ramesh2001a,Sasikumar2016}). We chose the monthly minimum values of the diameters since they are more likely to represent the `quiet' solar corona which at low radio frequencies is supposed to be the corona that is nearly free of localized thermal and/or non-thermal sources (see for e.g. \citeA{Leblanc1969a}).

\section{Analysis and Results}

\begin{figure}
\noindent\includegraphics[width=\textwidth]{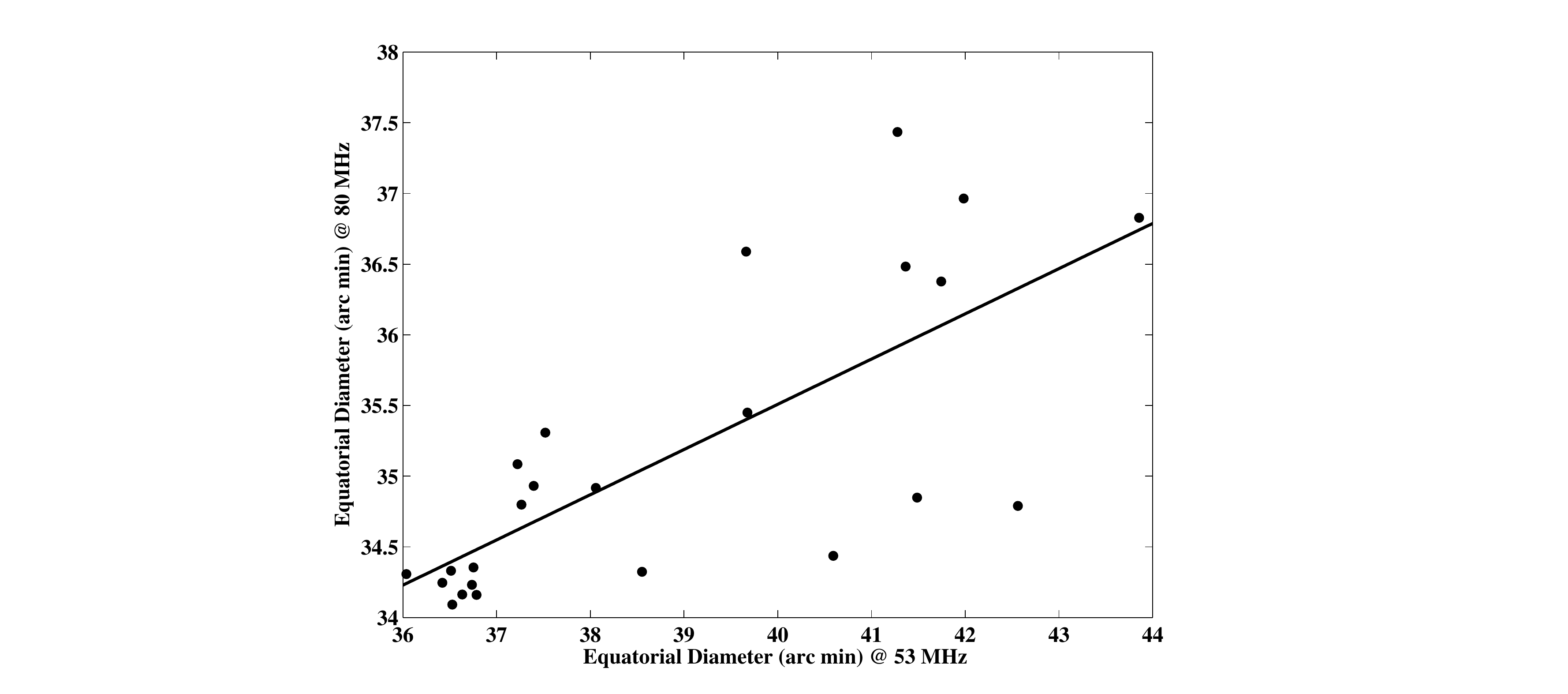}
\caption{Correlation between the 53 MHz and 80 MHz diameters in Figure \ref{figure3}. The correlation coefficient is 74\%. The `black' line is the linear least squares fit to the data points.}
\label{figure4}
\end{figure}

An inspection of Figure \ref{figure3} 
and the quadratic least squares fits to the data points there reveals a gradual decrease in the equatorial diameters at both 53 MHz and 80 MHz 
from January 2015. There is a high degree of correlation ($\approx$74\%) between the diameter values at the above two frequencies (Figure \ref{figure4}). This indicates that the discrete sources have been uniformally identified and removed at both 53 MHz and 80 MHz. Note that if the diameters had been over estimated due to possible incomplete removal of the discrete sources, particularly near the limb(s), then the distribution of the monthly minimum diameter values at 53 MHz and 80 MHz could have been random instead of exhibiting a decreasing trend as mentioned above. 
These are in support of the methodology described in Section 3 to estimate the diameters in the present case. 

The decrease in the diameter values at 53 MHz and 80 MHz in Figure \ref{figure3} is an indication of the decrease in the heliocentric distances ($r$) of the `critical' plasma levels corresponding to 53 MHz and 80 MHz in the solar atmosphere. It is well known that most of the observed thermal radio emission from the `quiet' Sun at any given radio frequency ($f$) orginates only from the corresponding `critical' level in the solar atmosphere. Since plasma frequency 
$f_{p}{\propto}\sqrt{N_{e}}$, and $f_{p}{=}f$ (where $f$ is the frequency of observation) at the critical `plasma' level mentioned above, the aforementioned decreases indicate that there should have been a corresponding decrease in $N_{e}$ during the same interval. The quadratic least squares fits to the data points in Figure \ref{figure3} indicate that the 
the 53 MHz equatorial diameter in 2019 January (${\approx}37^{\prime}$) is nearly same as the 80 MHz equatorial diameter in 2015 January. Since 
$N_{e}^{80}{\approx}7.9\times10^{7}\rm cm^{-3}$ and
$N_{e}^{53}{\approx}3.5\times10^{7}\rm cm^{-3}$, the above implies that 
$N_{e}$ at the corresponding heliocentric distance  
(i.e. ${\approx}$1.16$\rm R_{\odot}$) had decreased by a factor $\approx$2.3 between 2015 January and 2019 January. Comparing this with the published reports on $N_{e}$ in the equatorial region of the `quiet' Sun, we find that  $N_{e}{\approx}3.5\times10^{7}\rm cm^{-3}$ appears to the lowest value ever reported at $r{\approx}$1.16$\rm R_{\odot}$ (see for e.g. Figure 2 in \citeA{Warmuth2005}). 
This implies that radio emission at 53 MHz from the `quiet' Sun originated at $r{\approx}$1.16$\rm R_{\odot}$ during the minimum phase of sunspot cycle 24.
Note that the average height of the coronal type II radio bursts during the declining phase of sunspot cycle 23 was found to be smaller compared to their average height during the maximum phase due to a similar decrease 
in $N_{e}$ \cite{Gopalswamy2009}.

We verified the aforementioned decrease in $N_{e}$ during the descending phase of sunspot cycle 24 by independently estimating $N_{e}$ at $r{\approx}$1.55$\rm R_{\odot}$ in the equatorial region of the `quiet' corona
using the linearly polarized brightness (pB) measurements with 
STEREO-A/COR1 (http://sd-www.jhuapl.edu/secchi/lz/L0/a/seq/) and the inversion technique based on spherically symmetric polynomial approximation 
(SSPA; \citeA{Wang2014,Wang2017}; solar.physics.montana.edu/wangtj/sspa.tar).  
The $N_{e}$ estimates were obtained at $r{\approx}$1.55$\rm R_{\odot}$ since
it is the nearest distance to/above the STEREO-A/COR1 occulting disk 
(radius${\approx}$1.4$\rm R_{\odot}$) which is close to the equatorial diameters indicated by the overall fits in Figure \ref{figure3}.
A set of three STEREO-A/COR1 images taken sequentially around $\approx$06:30 UT with the linear polarizer at angles $0^{\circ}$, $120^{\circ}$, and $240^{\circ}$ with respect to the reference direction were used to construct the pB image. The time interval between the image obtained at each of the aforementioned positions of the polarizer is $\approx$9 sec. Note that STEREO-A/COR1 pB images obtained at a different time on the same day could have also been used since our interest is mainly in the `quiet' corona whose characteristics like $N_{e}$ generally vary on comparitively longer time scales only.
Note that there was no data available from STEREO spacecraft for 2015 January - 2015 December as it passed behind the Sun.
So we used 2016 January - 2019 May data. We analysed data obtained on three different days each month, with an interval of 10 days, during the above period. The images were processed to align with the solar North, and calibrated to remove the instrumental scattered light. The background was subtracted using data derived from a complete solar rotation.
We used SSPA to obtain 2D maps of $N_{e}$ along the equator, i.e. at position angles (PA) ${\approx}90^{\circ}$ and ${\approx}270^{\circ}$. We averaged the $N_{e}$ from 
PA ${\approx}90^{\circ}{\pm}15^{\circ}$ and ${\approx}270^{\circ}{\pm}15^{\circ}$ at $r{\approx}$1.55$\rm R_{\odot}$ to reduce the variation in $N_{e}$.

\begin{figure}
\noindent\includegraphics[width=\textwidth]{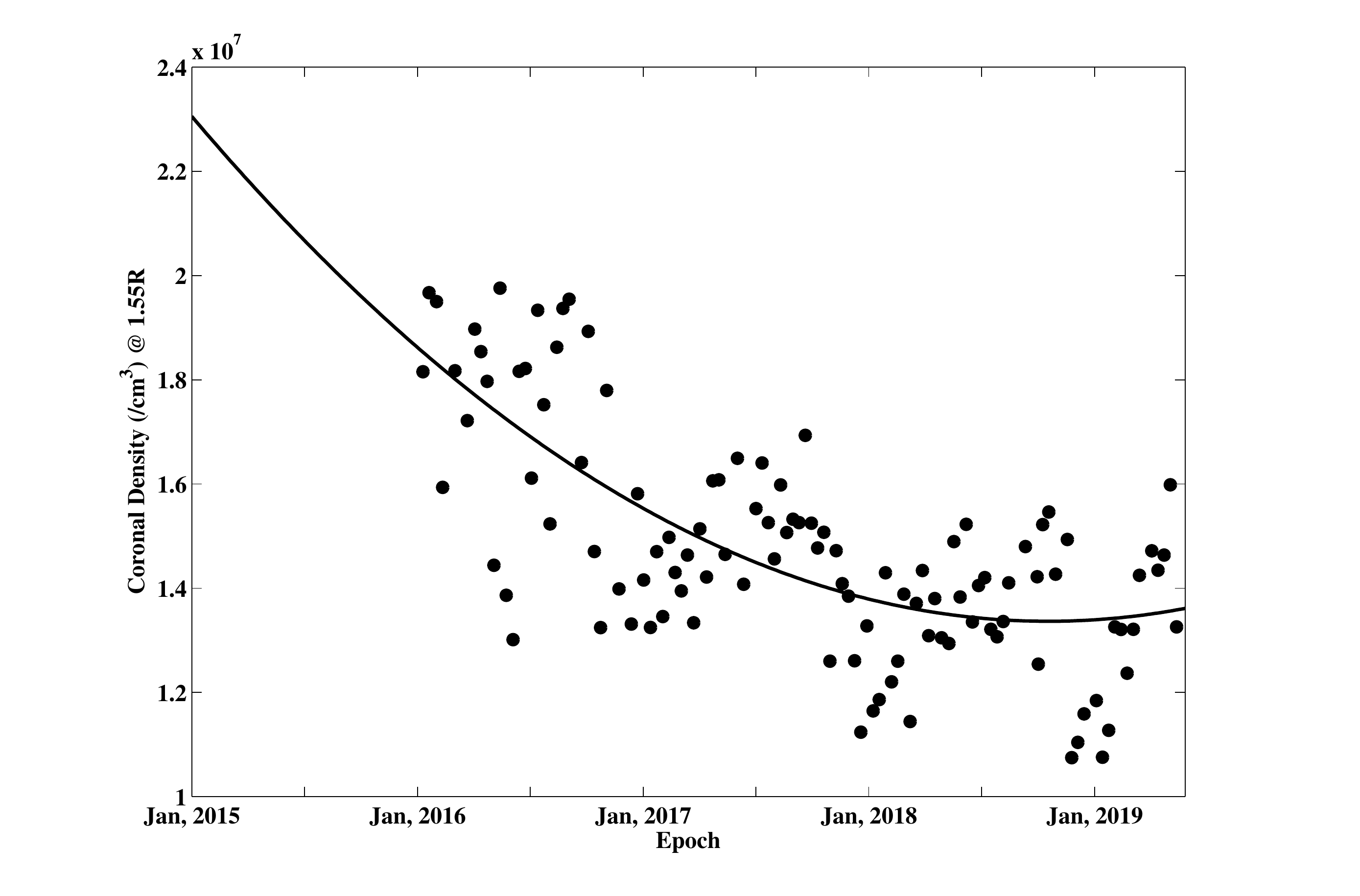}
\caption{Estimates of $N_{e}$ at $r{\approx}\rm 1.55R_{\odot}$ from white-light (STEREO-A/COR1) observations during the period 2016 January - 2019 May. The `black' line is the quadratic least squares fit to the data points.} 
\label{figure5}
\end{figure}

Figure \ref{figure5} shows the $N_{e}$ values thus obtained on 117 different days during interval 2016 January - 2019 May when there were no coronal streamers and/or coronal holes in the above PA range. The scatter in the $N_{e}$ estimates are mostly due to the 
assumption of spherically symmetric approximation for the inversion and the errors associated with instrumental background subtraction due to bad pixels in the detector. 
The quadratic least squares fit to the data points in Figure \ref{figure5} indicates a decreasing trend in $N_{e}$. 
The maximum and minimum values of $N_{e}$ obtained from the fit 
are ${\approx}2.3{\times}10^{7}\rm cm^{-3}$ (2015 January) 
and ${\approx}1.3{\times}10^{7}\rm cm^{-3}$ (2019 January),
respectively. This implies that $N_{e}$ had decreased by a factor of ${\approx}$1.8 at 
$r{\approx}$1.55$\rm R_{\odot}$ in the above interval during the descending phase of sunspot cycle 24. 
This is reasonably consistent with the ${\approx}2.3\times$ decrease in $N_{e}$ estimated from the GRAPH observations at $r{\approx}$1.16$\rm R_{\odot}$ 
(see Figure \ref{figure3}).
The nearly equal factor by which $N_{e}$ has decreased at two different heliocentric distances suggest an overall decline in $N_{e}$. 

\section{Summary}

We investigated the changes in the size of the solar corona during the period 2015 January - 2019 May (descending phase of sunspot cycle 24) using data obtained with the GRAPH at 53 MHz and 80 MHz. Our results indicate that the heliocentric distances at which radio emission at the above two frequencies orginate in the equatorial region of the `quiet' solar corona gradually decreased during the said period
due to a decrease in $N_{e}$. We independently verified the latter using white-light coronagraph estimates of $N_{e}$ in the equatorial region of `background' corona that was devoid of coronal streamers and/or coronal holes. 

\acknowledgments
We are indebted to the staff of the Gauribidanur observatory for their
help in maintenance of the antenna/receiver systems, and the observations.
The STEREO/SECCHI data used here are produced by an international consortium of the Naval Research Laboratory (USA), Lockheed Martin Solar and Astrophysics Lab (USA), NASA Goddard Space Flight Center (USA), Rutherford Appleton Laboratory (UK), University of Birmingham (UK), 
Max-Planck-Institut f\"{u}r Sonnensystemforschung (Germany), Centre Spatiale 
de Li\'{e}ge (Belgium),  Institut d{'}Optique Th\'{e}orique et Appliqu\'{e}e (France), 
and Institut d{'}Astrophysique Spatiale (France).
The white-light $N_{e}$ values reported are due to the code developed by T.J.Wang for STEREO-COR1 pB observations. 
The SOHO data are produced by a consortium of the Naval Research Laboratory (USA),
Max-Planck-Institut fuer Aeronomie (Germany), Laboratoire
d'Astronomie (France), and the University of Birmingham (UK). SOHO
is a project of international cooperation between ESA and NASA.
One of the authors (AK) acknowledges the European Research Council (ERC) under the European Union's Horizon 2020 Research and Innovation Programme Project SolMAG 724391. A. Kumari, D. Ketaki, and M. Vrunda were at the Gauribidanur observatory when the work was carried out. Data used in the study are available at https://www.iiap.res.in/gauribidanur/home.html. We thank the referees for their comments which helped us to present the results in a better manner.


\end{document}